\documentclass[
    a4paper,
    amsfonts,
    amssymb,
    amsmath,
    reprint,
    showkeys,
    nofootinbib,
    twoside,aps,pra
]{revtex4-2}

\usepackage[english]{babel}
\usepackage[utf8]{inputenc}
\usepackage{booktabs}           
\usepackage[table,xcdraw]{xcolor} 

\usepackage[colorinlistoftodos, color=green!40, prependcaption]{todonotes}

\usepackage[
    pdftex,
    pdftitle={Article},
    pdfauthor={Author}
]{hyperref}
\expandafter\let\csname equation*\endcsname\relax
\expandafter\let\csname endequation*\endcsname\relax
\usepackage{amsmath}
\usepackage{overpic}
\usepackage{graphicx}
\usepackage{subcaption}
\usepackage{braket}
\usepackage{orcidlink}
\usepackage{xcolor}
\definecolor{dark blue}{rgb}{0.0,0.0,0.55}
\graphicspath{/figures}

\newcommand{\average}[1]{\left\langle#1\right\rangle}
\usepackage{amsmath}
\usepackage{cleveref}

\begin{document}

\preprint{AAPM/123-QED} 

\title[Sample Title]{Fast and efficient long-distance quantum state transfer in long-range spin-$\frac{1}{2}$ models}

\author{F. Faria$^1$\orcidlink{0009-0007-0457-9514}, C.C. Nelmes$^{1*}$\orcidlink{0009-0002-1686-6282}, T.J.G. Apollaro$^2$\orcidlink{0000-0002-9324-9336}, T.P. Spiller$^1$,$^3$\orcidlink{0000-0003-1083-2604}, and I. D'Amico$^1$,$^3$\orcidlink{0000-0002-4794-1348}}

\affiliation{$^1$School of Physics, Engineering and Technology, University of York, York, YO10 5DD, United Kingdom}

\affiliation{$^2$Department of Physics, University of Malta, Msida MSD 2080, Malta}

\affiliation{$^3$York Center for Quantum Technologies, University of York, York, YO10 5DD, United Kingdom}

\affiliation{$^*$Author to whom any correspondence should be addressed.}

\email{Correspondence email: c.nelmes@york.ac.uk}

\date{\today}


\begin{abstract}
Quantum state transfer is investigated beyond the nearest-neighbour coupling scheme in long spin-$\frac{1}{2}$ linear chains. Exploiting the properties of the next-nearest neighbour Hamiltonian's dispersion relation, it is shown that with minimal engineering, i.e., an on-site magnetic field on the two end sites and only a few symmetrically-modified end inter-site couplings, an average transfer fidelity above $99\%$ can be achieved.  To leading order, the required time scales linearly with the length of the chain.  Such a fast, high-quality quantum state transfer is based on the ballistic propagation of the wave packet centred in the linear region of the dispersion relation by means of the on-site magnetic field. At the same time, the wave packet width, modulated by the inter-site couplings at the chain ends, whose values are found via a carefully designed genetic algorithm, is constrained mostly in the linear region of the dispersion relation. Our coupling scheme is shown to hold for arbitrary values of the next-nearest inter-site coupling and can be straightforwardly applied to longer range coupling schemes.
\end{abstract}

\maketitle

\section{Introduction}
The transfer of quantum information between different locations is paramount for a variety of quantum information processing tasks, ranging from quantum key distribution to quantum computation~\cite{Nielsen2010}. In a seminal paper~\cite{Bose1}, Bose introduced a quantum state transfer (QST) protocol based on the coherent dynamics of a quantum channel,  modelled by a spin-$\frac{1}{2}$ chain. Such a protocol does not require the physical transfer of the carrier of the quantum information between the sender's and receiver's location, as performed, for example, by employing `flying qubits', such as photons~\cite{Northup2014,Wei2015,Chan2022}.

Since Bose's paper, where it was shown that the QST fidelity via uniformly-coupled Heisenberg spin chains quickly decays below the classical limit of $\frac{2}{3}$ by increasing the chain's length, intense theoretical and experimental investigations for fast and efficient long-distance QST have taken place \cite{chapman2016,Li2018,Tian2020,Roy2025}. In Refs.~\cite{Karbach2005,Christandl1,Nikolopoulos_2004,Kay2011,PhysRevLett.101.230502,Wang2011a,PhysRevA.74.030303,vinetHowConstructSpin2012} several coupling schemes attaining perfect state transfer (PST) at the relevant quantum speed limit have been introduced. These coupling schemes generally involve modifying all of the nearest-neighbor couplings along the linear chain. Although PST is certainly a desirable primitive, both fully-engineering all the couplings and unavoidable fabrication and read-out errors may spoil the efficacy of this PST scheme. Hence, less demanding coupling schemes have been introduced. It has been realised that near-perfect, fast QST can be achieved by modifying symmetrically only a few couplings at each end of the spin chain~\cite{TA2012,Banchi_2011,PhysRevA.85.012318}. The mechanism behind this QST dynamics is ballistic wave packet propagation, which is enabled by just a few modified couplings being sufficient to concentrate the wave packet in $k$-space in the linear region of the energy spectrum, hence minimising detrimental dispersion effects. Other schemes for QST utilize, e.g.: chains where the sender and receiver spin are perturbatively coupled to the quantum channel~\cite{wojcikUnmodulatedSpinChains2005,Lorenzo2013,almeidaQuantumstateTransferStaggered2015}, which, in general, produce long QST times; dimerized spin chains with topological defects \cite{Estarellas_2017,Wilkinson2017}, which can be effective beyond the perturbative coupling regime for sender and receiver; modulation of on-site energies instead of couplings \cite{B2025,nelmes2025statetransferanalysislinear}; engineering of constructive quantum state interference \cite{Karbach2005,Christandl2} often characterised by a complicated quantum state dynamics and relatively long QST times. 

Whereas the majority of QST coupling schemes focus on spin chains with only nearest-neighbour (NN) interactions, it is of great interest to investigate QST beyond the NN scenario. On the one hand, long-range spin chains model a great variety of physical systems~\cite{RevModPhys.95.035002}, on the other hand, it can be expected that, by extending the range of the interactions, faster QST (with respect to the NN scenario) can be achieved~\cite{PhysRevA.102.010401}. Among the above reported QST coupling schemes, both the PST~\cite{Kay2006,Niko1,Christandl2} and the perturbatively-coupled~\cite{Gualdi2008,Ronke_2011,hermesDimensionalityenhancedQuantumState2020} approaches have been investigated beyond the NN scenario. However, in neither coupling scheme is a reduction of the QST time achieved. For the NNN PST scheme, the (non-normalised) transfer time $T$ is equivalent to that of the NN scenario and for the perturbatively-coupled scheme, the transfer time is generally very long, being of the order of the inverse of the separation of quasi-degenerate energy eigenstates.  A recent work has shown that the extended $XY$ model, involving $N$-body interactions that decay with the distance, does indeed provide faster QST although at the price of significantly reduced fidelity~\cite{Ahuja}.
Interestingly, it has not yet been fully investigated whether long-range interacting spin-$\frac{1}{2}$ models are capable of fast, high-quality QST utilising ballistic dynamics.

In our paper, we show that ballistic, fast and high-quality long-distance QST is possible in long-range interacting spin chains. Building on the optimal end-coupling scheme proposed in Refs.~\cite{TA2012,Banchi_2011}, 
we show that applying an additional on-site energy term on the end sites allows us to achieve a quasi-dispersionless wave packet dynamics. We attain an average fidelity greater than $99\%$ in times significantly shorter than those required by the NN scenario. Our analysis, which is grounded on tight analytic approximations for the wave packet dynamics, allows us also to provide a clear physical picture of the QST scheme and to guide the optimisation procedure of the end couplings in an efficient way via a genetic algorithm.

The paper is organised as follows: in Sec.~\ref{sec2} we introduce the NNN spin-$\frac{1}{2}$ linear chain, the relevant QST's figure of merit and the analytic approximation about the model's energy spectrum, i.e., its dispersion relation; in Sec.~\ref{sec3} we introduce the genetic algorithm and focus on performing the optimisation procedure on a specific parameter instance of the NNN model; in Sec.~\ref{sec4} we present our results, showing that fast, high-quality QST is achieved by optimising a limited number of end couplings and provide a thorough analysis of the achieved result. Finally, in Sec.~\ref{conc} we draw our conclusions.

\section{Spin Model}\label{sec2}

\begin{figure}[h!]
    \includegraphics[width=\linewidth]{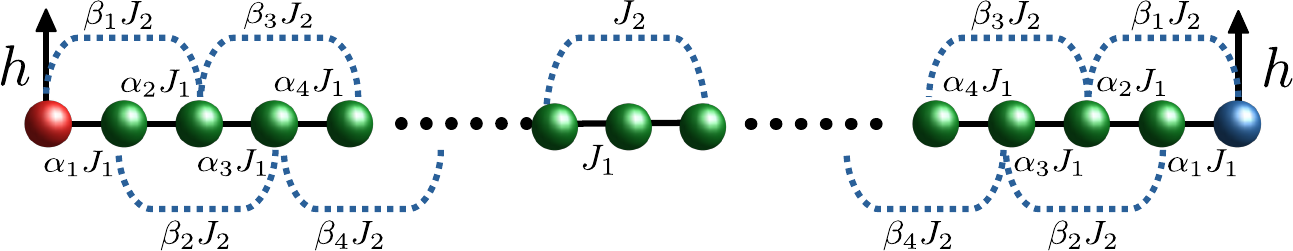}
    \caption{Next-nearest neighbour spin chain with uniform bulk couplings $J_1$ and $J_2$. End site couplings are symmetrical rescaled by $\{\alpha_i\}$ for NN couplings and $\{\beta_i\}$ for NNN couplings. In the figure $i=1,2,3,4$ as our results show that modifying the couplings of just four end sites is sufficient for fast and high-quality QST for long chains. The black arrow on the first and last site represent equal transverse applied magnetic fields, which generate an on-site energy $h$. The sender and the receiver qubit, respectively the red and the blue sphere, are located at the ends.}
    \label{fig:spinchaindiagram}
\end{figure}

We consider an $XX$ spin-$\frac{1}{2}$ chain described by the Hamiltonian 
\begin{multline}\label{eq:2NNXYHamiltonian}
\hat{H}=\frac{J_1}{4}\sum_{i=1}^{N-1}\alpha_i\left(\hat{\sigma}^x_{i}\hat{\sigma}^x_{i+1}+\hat{\sigma}^y_{i}\hat{\sigma}^y_{i+1}\right)\\+\frac{J_2}{4}\sum_{i=1}^{N-2}\beta_i\left(\hat{\sigma}^x_{i}\hat{\sigma}^x_{i+2}+\hat{\sigma}^y_{i}\hat{\sigma}^y_{i+2}\right)+\frac{h}{2}\left(\hat{\sigma}_1^z+\hat{\sigma}_N^z\right)~,
\end{multline}
where $\hat{\sigma}_i^{x,y,z}$ are the usual Pauli matrices for the spin located at site $i$, and $J_1$ and $J_2$ are, respectively, the nearest-neighbour and the next-nearest neighbour couplings and the on-site energy $h$ represents the local magnetic field. The set of values $\{\alpha_i\}$ and $\{\beta_i\}$ are rescaling factors for the NN and NNN couplings, respectively, of the spin sitting on site $i$, where $\alpha_i=\alpha_{N-i}$ and $\beta_i=\beta_{N-1-i}$ for mirror-symmetry. See Fig.~\ref{fig:spinchaindiagram} for a schematic diagram of the model. 
In the following, we set $J_1=1$ as our energy and inverse time unit and $0\leq J_2 \leq 1$ as, in general, the inter-site coupling strength decays with the distance. $J_1=1$ corresponds also to our maximum coupling $J_{max}$ in the system. 
The Hamiltonian in Eq.~\ref{eq:2NNXYHamiltonian} possesses $U(1)$ symmetry and, hence, the total magnetisation along the $z$-axis is conserved. This allows us to express Eq.~\ref{eq:2NNXYHamiltonian} as a direct sum of subspaces each with a fixed number $n$ of excitations (spin-down states), $\hat{H}=\bigoplus_{n=0}^N\hat{H}_n$.

The QST scheme we investigate has been proposed by Bose in Ref.~\cite{Bose1} and entails that an unknown single-qubit quantum state $\ket{\psi_s}=\cos\frac{\vartheta}{2}\ket{0}+\sin\frac{\vartheta}{2}e^{i \phi}\ket{1}$ is encoded at some location (the sender site) and, exploiting the unitary dynamics of the spin chain, retrieved at time $t$ at some different location (the receiver site). Hereafter, we set the sender and the receiver site at the opposite ends of the chain, i.e, site 1 and site $N$. The quality of the QST protocol is hence assessed by evaluating the fidelity between the sender and the receiver state, averaged over all pure input states,
\begin{align}
    \label{eq:averageF}
    \average{F(t)}=\frac{1}{4\pi}\int d\Omega \bra{\Psi_s}\hat{\rho}_r(t)\ket{\Psi_s}~.
\end{align}
Initialising all but the sender spin in the fully polarised state, i.e., $\ket{\Psi(0)}=\ket{\psi}\otimes_{i=2}^{N}\ket{0}_i$, we can exploit the model's $U(1)$-symmetry and restrict the dynamics to the zero- and one-particle subspace, yielding for Eq.~\ref{eq:averageF} \cite{Bose1}
\begin{align}
    \label{eq:averageF1}
    \average{F(t)}=\frac{1}{2}+\frac{\left|f_1^N(t)\right|}{3}+\frac{\left|f_1^N(t)\right|^2}{6}~,
\end{align}
where $f_1^N(t)=\bra{N}e^{-i \hat{H}_1 t}\ket{1}$ is the single-particle transition amplitude from site 1 to site $N$. Hence, the average fidelity $\average{F(t)}$ is a monotonically increasing function of the absolute value of the transition amplitude, which is given by
\begin{align}
\label{eq:trans1}
    f_1^N(t)=\sum_{k=1}^N v_{1k}v_{Nk}^*e^{-i \omega_k t}~,
\end{align}
where $\{\omega_k, \ket{v_k}\}$, with $\ket{v_k}=\ket{v_{1k},v_{2k},\cdots,v_{Nk}}$ are the eigenvalues and eigenvectors of $\hat{H}_1$, i.e., the Hamiltonian in the single-particle sector.
Finally, exploiting the fact that the Hamiltonian is real and symmetric both with respect to the main diagonal and the skew diagonal (mirror-symmetry), the first and last component of the $k$ eigenvectors satisfy $v_{1k}=(-1)^k v_{Nk}$~\cite{Cantoni1976} and Eq.~\ref{eq:trans1} can be cast as
\begin{align}
\label{eq:trans2}
    f_1^N(t)=\sum_{k=1}^N v_{1k}^2 e^{-i \left(\omega_k t-k\pi\right)}~.
\end{align}
As a result, Eq.~\ref{eq:trans2} can be interpreted as a wave packet with $k$ components evolving according to the dispersion relation $\omega_k$. Following Refs.~\cite{TA2012,Banchi_2011}, we aim at achieving high-quality QST by  concentrating the initial wave packet $v_{1k}^2$ into a mainly linear region of the dispersion relation $\omega_k$, so that quasi-dispersionless ballistic excitation transfer is realised.

Let us start the analysis of Eq.~\ref{eq:trans2} with the dispersion relation $\omega_k$ arising from the uniform model in Fig.~\ref{fig:spinchaindiagram}, i.e., $\alpha_i=\beta_i=2$, for all $i$ and $h=0$. In this case $\omega_k$
is given by the eigenvalues of the $N\times N$ pentadiagonal symmetric Toeplitz matrix with elements $T_{ij}=J_1\left(\delta_{i,j+1}+\delta_{i,j-1}\right)+J_2\left(\delta_{i,j+2}+\delta_{i,j-2}\right)$. Although the eigenvalues $\omega_k$ can be expressed as zeros of rational functions~\cite{Elouafi2011}, here we will follow a different approach which will prove itself useful in deriving analytical approximations of the properties of the dispersion relation $\omega_k$. This will allow us to formulate educated estimates for the values of the on-site end couplings for high-quality QST and, at the same time, provide an intuitive picture of the underlying dynamics.

Banded Toeplitz matrices are asymptotically equivalent~\cite{Zhu2017}, with a upper-bounded error of $O(1/N)$, to circulant matrices, whose eigenvalues are well-known~\cite{Gray2001}. In other words, we approximate the eigenvalues of our open-boundary spin chain with those of a periodic-boundary model. Clearly, the longer the chain and the lower the order $m+1$ of the banded Toeplitz matrix (in our case $m=2$), the better the approximation. The eigenvalues of the circulant matrix are given by
\begin{equation}\label{eq:oddcirculanteigs}
    \omega_k =  J_1 \cos \left (\frac{2 \pi k}{N}\right )+ J_2 \cos \left (\frac{4 \pi k}{N}\right )~,~k=1, 2,\dots,N~,
\end{equation}
where, for $N\rightarrow \infty$,
\begin{equation}\label{eq:oddcirculanteigs2}
    \omega\left(\theta\right) =  J_1 \cos \theta+ J_2 \cos 2\theta~,~\theta\in \left[0,2\pi\right]~.
\end{equation}
Finally, because of the open boundary conditions of our model, the allowed values of $\theta$ in Eq.~\ref{eq:oddcirculanteigs2} are restricted to the interval $\left[0,\pi\right]$.

 \begin{figure}
     \includegraphics[width=1\linewidth]{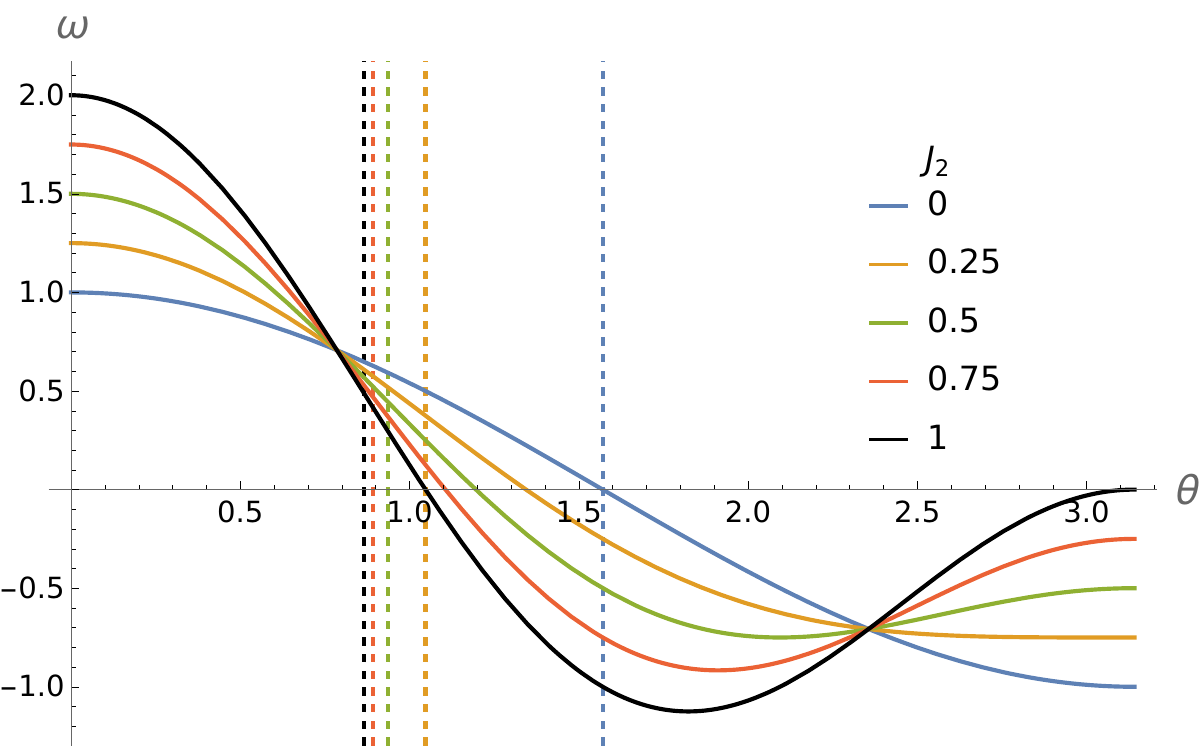}
     \caption{Dispersion relation, Eq.~\ref{eq:oddcirculanteigs2}, for different values of $J_2$, as labelled. The vertical dashed lines indicate the inflection point $\theta_1$ in Eq.~\ref{eq_inflection_points} for each corresponding NNN coupling. By increasing $J_2$, the linear region of the dispersion relation shifts towards higher energies $\omega$ and, at the same time, the group velocity $\frac{d\omega}{d\theta}$ increases.}
     \label{fig:dispersion_relation}
 \end{figure}

In Fig.~\ref{fig:dispersion_relation}, we plot the dispersion relation in Eq.~\ref{eq:oddcirculanteigs2} for different values of $J_2$. It is evident that, with respect to the nearest-neighbor case $J_2=0$, the dispersion relation acquires a more complex structure: a.) the inflection point, $\frac{d^2\omega}{d\theta^2}=0$, around which the linear region extends, is found at higher energies, while, at the same time, the extension of the linear region shrinks; b.) there is a threshold value of $J_2$ where two inflection points, and hence two linear regions of $\omega$, are found; c.) the group velocity, $\frac{d\omega}{d\theta}$, of the wave packet centred at the higher-energy inflection point increases by increasing $J_2$. These properties, hence, inspire the physical mechanism for fast and efficient QST in the NNN model: apply an on-site magnetic field on the first and last site in order to shift the peak of the wave packet to the inflection point and modify only a limited number of end couplings in order to concentrate the modes entering the transition amplitude, Eq.~\ref{eq:trans2}, in the reduced linear region.
Furthermore, the circulant approximation also allows us to determine the value of the on-site energies, as well as the potential speed-up of the QST protocol in the NNN model. 
The single-particle spectrum of Eq.~\ref{eq:2NNXYHamiltonian} is bounded by
\begin{equation}
    \left[\omega_{min},\omega_{max}\right]=
    \begin{cases}
        \left[-J_1+J_2,J_1+J_2\right]~,~&0\leq J_2 \leq \frac{1}{4}\\
        \left[-\frac{J_1^2}{8 J_2}-J_2,J_1+J_2\right]~,~&\frac{1}{4} \leq J_2 \leq 1
    \end{cases}~,
\end{equation}
while the inflection points are located at
\begin{eqnarray}\label{eq_inflection_points}
    \begin{cases}
        \theta_1=\cos^{-1}\left(-\frac{J_1}{16 J_2}+\sqrt{\left(\frac{J_1}{16 J_2}\right)^2+\frac{1}{2}} \right)\\
        \theta_2=\pi-\cos^{-1}\left(\frac{J_1}{16 J_2}+\sqrt{\left(\frac{J_1}{16 J_2}\right)^2+\frac{1}{2}} \right)
    \end{cases}~,
\end{eqnarray}
with $\theta_1$ for $0\leq J_2 \leq \frac{1}{4}$ and $\left\{ \theta_1,\theta_2\right\}$ for $\frac{1}{4}\leq J_2 \leq 1$.
Hence, an educated estimate for the local magnetic fields on site 1 and $N$ can be made
\begin{eqnarray}\label{eq_hvalues}
  h=2 \omega(\theta_1)=-\frac{3 J_1\left(\frac{J_1}{J_2}-\sqrt{128+\left(\frac{J_1}{J_2}\right)^2} \right)}{32}~.
\end{eqnarray}
As for the maximum speed-up of the QST protocol in the NNN model with respect to the NN case, we
determine the ratio $s$ of their group velocities for a wave packet centred at their respective inflection points,
\begin{multline}\label{eq:speedup}
    s=\frac{\frac{d\omega_{NNN}}{d\theta}\bigg|_{\theta=\theta_1}}{\frac{d\omega_{NN}}{d\theta}\bigg|_{\theta=\frac{\pi}{2}}}\\=\scalebox{0.9}{$\frac{1}{32} \left(3 + \sqrt{1 + 128 \left(\frac{J_2}{J_1}\right)^2} \right)\sqrt{32 + \frac{J_1 \left( -J_1 + \sqrt{J_1^2 + 128 J_2^2} \right)}{2 J_2^2}}$}~.
\end{multline}
This speed-up $s$ is an upper bound and in Sec.~\ref{sec3} we will show that our proposed end-modified QST scheme is very close to it.
With the above analytic results, in the following sections we will introduce the numerical methods utilised for determining the values of the end couplings for the QST in the NNN model. Without loss of generality, we will showcase the results for $J_2=0.5$, where the ratio $s\simeq1.76$.

\section{Genetic Algorithm}\label{sec3}
\subsection{Background}
Emerging as a subset of evolutionary computation, genetic algorithms have proven to be an effective metaheuristic for optimisation problems \cite{H1992,E2015,L2019,Feng2019}. The algorithmic process typically begins with initiating a population of randomised individuals which are then assessed against a pre-existing fitness function. Individuals within this population that score higher, with respect to the fitness function, actively steer the algorithm towards traits that allow for the highest fitness scores. Once the fitness scores are recorded, a crossover function exchanges the `genes' amongst the higher scoring individuals in a way reminiscent of biological Darwinian evolution. A mutation process then alters the individuals in a way that is either random or in an attempt to preserve certain traits that are understood as necessary. 

Genetic algorithms have been used specifically as a powerful tool in optimising parameters within spin chains and networks for ideal state transfer entanglement generation and transfer, and quantum information processing \cite{B2025,LM2021,PS2025}. The parameters which are conventionally optimised within these kinds of system are the matrix elements of the  spin Hamiltonian, which represent the inter-qubit couplings and site-specific qubit energies. If high-fidelity and rapid state transfer is desirable within a specific system, then the couplings and on-site energies may be optimised within the genetic algorithm to allow for this kind of transfer.
\subsection{Execution}
The genetic algorithm used within this study stems from that which was used within Refs.~\cite{B2025,LM2021}. It maintains the algorithmic parameters shown in Table. \ref{params}, as well as the crossover function and mutation process (barring a few minor alterations). Notably, there is an alteration of the fitness function  to
\begin{eqnarray}
x(f_1^N(\tau),\tau) = 100\text{e}^{(a(|f_1^N(\tau)|^2-1))},
\label{ff}
\end{eqnarray}
which tracks the maximum excitation transfer fidelity attained at time $\tau$ (in units of $1/J_{max}$), within the time interval of $t\cdot J_{max}\in[0,N]$. The parameter $a$ is a  constant set to 10 consistent with the previous study. The change of the mutation process was also enforced such that the final genome, containing the list of couplings, is `mirrored' about its middle to reduce the search space to individuals which yield the necessary mirror symmetry for quasi-perfect/perfect state transfer \cite{B2025}.
\begin{table}[h!]
\centering
\small
\begin{tabular}{|l|c|c|}
\hline
\textbf{Generations} & \(\mathbf{{Population-Size}}\) & \(\textbf{Initial Mutation-Rate}\)  \\ \hline
\hspace{0.6cm} 200 & 1026 & ${20\%}$\\ \hline
\end{tabular}
\caption{Tabled values of the algorithmic parameters used for the genetic algorithm which discovered the numerical results shown within Sec.~\ref{sec3}. The initial mutation rate begins, as shown, at 20\% and is set to decrease as a function of the generations.}
\label{params}
\end{table}
The algorithm commences via an initialisation of a population of individuals of NNN chains where the bulk of the chain follows a pattern of NN coupling to NNN coupling ratio of $J_{2}/J_{1} = 1/2$. Dependent upon the number of sites that are being optimised (as there is a minimum of two couplings per site), the algorithm initialises a random configuration beginning from the outer sites. Therefore, if there is only one site being optimised from the beginning of the chain, then two couplings (the NN between the first to the second site and the NNN between the first to the third) are randomly generated and then subjected to the fitness evaluation via the fitness function in Eq.~\ref{ff}. As mirror symmetry is enforced throughout the mutation process, the identical coupling profile is imposed on the opposite side of the chain. The local magnetic field strengths located at the first and last sites were also concurrently and simultaneously optimised.  

Once the fitnesses have been evaluated and recorded, a crossover function exchanges the genetic encoding of the parents with equal probability to the offspring. This process allows for the traits which might facilitate the highest-fidelity transfer to be passed on to the next generation of prospective solutions. Iterations of this process, over a sufficiently large population and number of generations (See Table. \ref{params}) allow for convergence to high-performing solutions. The fine tuning of the mutation rate, particularly one that decreases over generations, is advantageous for finding more precise regions of local optima. 

\section{Fast and efficient QST}
\label{sec4}
In this Section we show that fast and efficient QST can be achieved in the NNN model by modifying symmetrically just a limited number of couplings at each end of the spin chain and, crucially, by adding an on-site energy on the first and last site. Without loss of generality, we focus on the case of chain's length $N=51$ and report in Table~\ref{tab:differentN} and Fig.~\ref{fig:logNvstimeandcoupling} the results for longer chains. 

\begin{figure}
    \centering
\includegraphics[width=\linewidth]{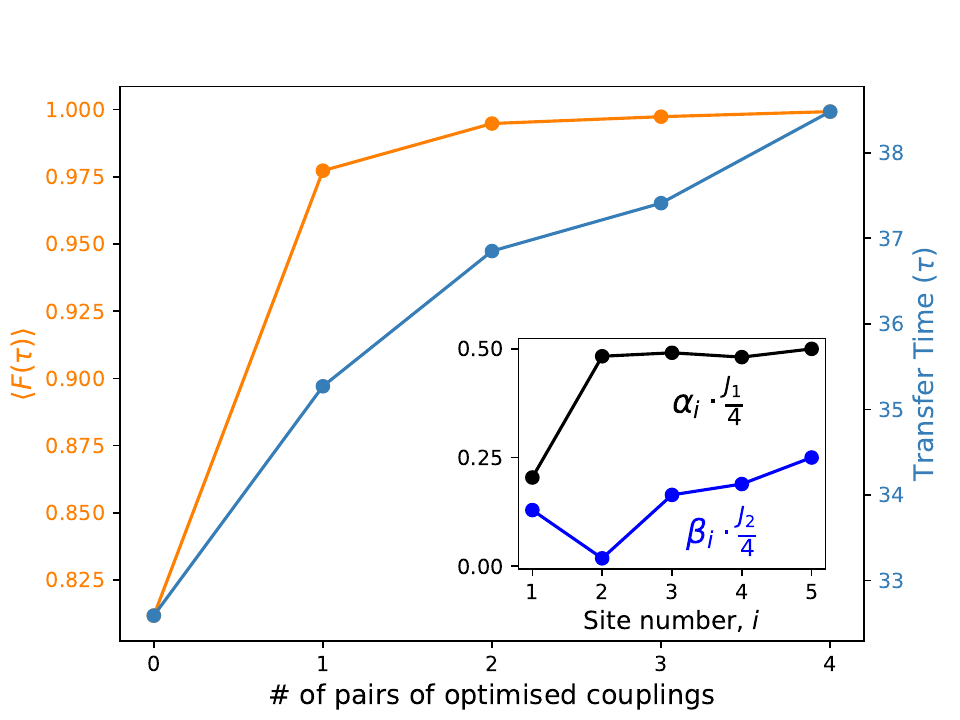}
    \caption{Average Fidelity (orange) and transfer time  (blue), in units of ($t \cdot J_{max}$), against the number of pairs of optimised sites for a 51 site spin chain. Inset shows the nearest neighbour (black), $\alpha_i \cdot \frac{J_1}{4}$, and next-nearest neighbour (blue), $\beta_i \cdot \frac{J_2}{4}$, coupling configuration for four pairs of optimised sites at one end (mirrored at the other end) with i = 5 corresponding to the unmodified bulk coupling value.}
    \label{fig:avgfid+timevsnumoptcouplingsj20.5}
\end{figure}

In the main panel of Fig.~\ref{fig:avgfid+timevsnumoptcouplingsj20.5} we report the maximum value of the average fidelity $\average{F(\tau)}$, Eq.~\ref{eq:averageF1}, for $\tau$ within the time interval $0\leq t\leq N$.  $\average{F(\tau)}$ increases monotonically to close-to-unity values by enlarging the set of optimised end couplings from one to four pairs $\{\alpha_i,\beta_i\}$, in addition to the on-site energy $h$.
Fig.~\ref{fig:avgfid+timevsnumoptcouplingsj20.5} shows also the QST transfer time $\tau$ which, while increasing by enlarging the set of optimised couplings, is significantly lower than the corresponding time for the nearest-neighbour coupling model, the latter scaling as $\tau\simeq N+ 3.2 N^{\frac{1}{3}}$~\cite{TA2012}.
In the inset of Fig.~\ref{fig:avgfid+timevsnumoptcouplingsj20.5} the values of $\{\alpha_i,\beta_i\}$ are shown for the case of four optimised pairs, i.e., up to $i=4$ in Eq.~\ref{eq:2NNXYHamiltonian}.
We observe that the genetic algorithm finds an almost monotonic increase in the coupling strength for the nearest-neighbour couplings $\alpha_i$, while the next-nearest neighbour couplings $\beta_i$ possess a near-zero coupling for $J_{2,4}$, i.e., $\beta_2\ll 1$. These trends are similarly observed for smaller numbers of optimised couplings.

\begin{figure}[h!]
    \centering

    \begin{minipage}[b]{.78\linewidth}
    \centering
    \begin{overpic}[width=\linewidth]{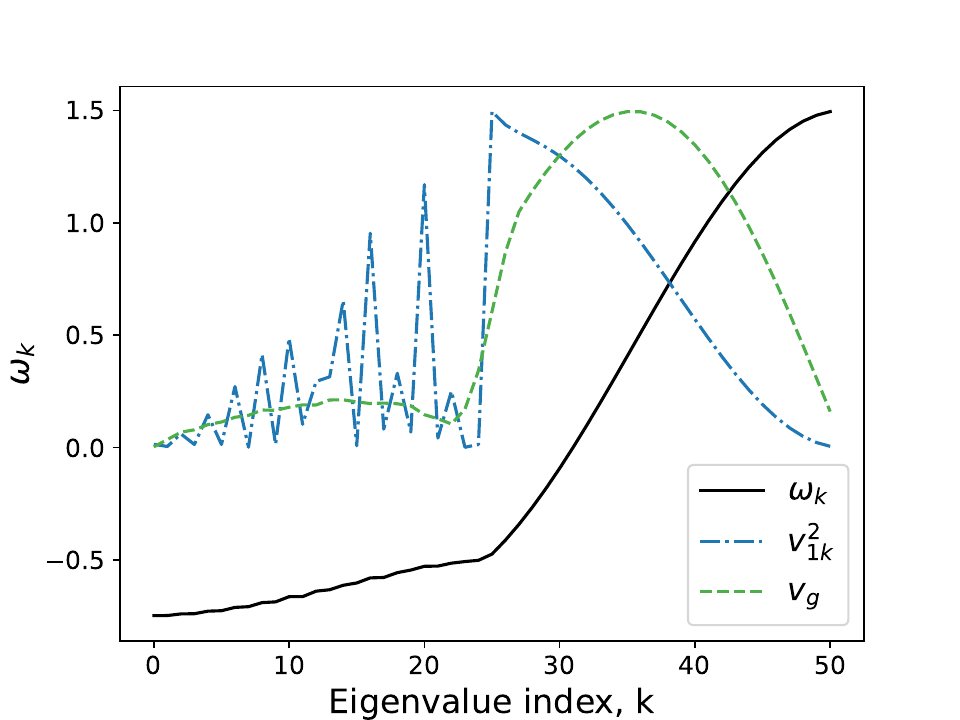}
        \put(15,60){\textbf{(a)}}
    \end{overpic}
    \end{minipage}\\[-0.75cm]
    \begin{minipage}[b]{.775\linewidth}
    \begin{overpic}[width=\linewidth]{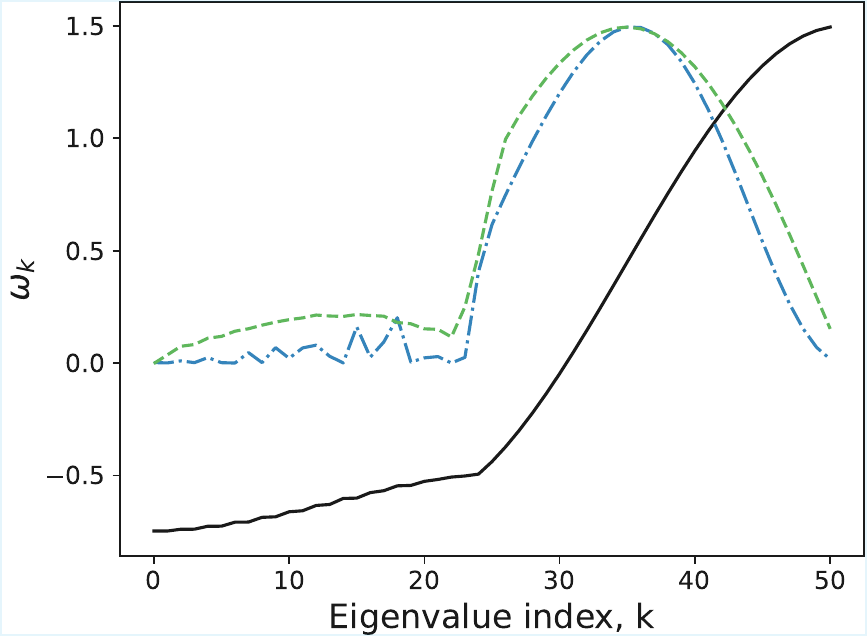}
        \put(15,60){\textbf{(b)}}
    \end{overpic}
    \end{minipage}\\[-0.75cm]

    \begin{minipage}[b]{.78\linewidth}
    \begin{overpic}[width=\linewidth]{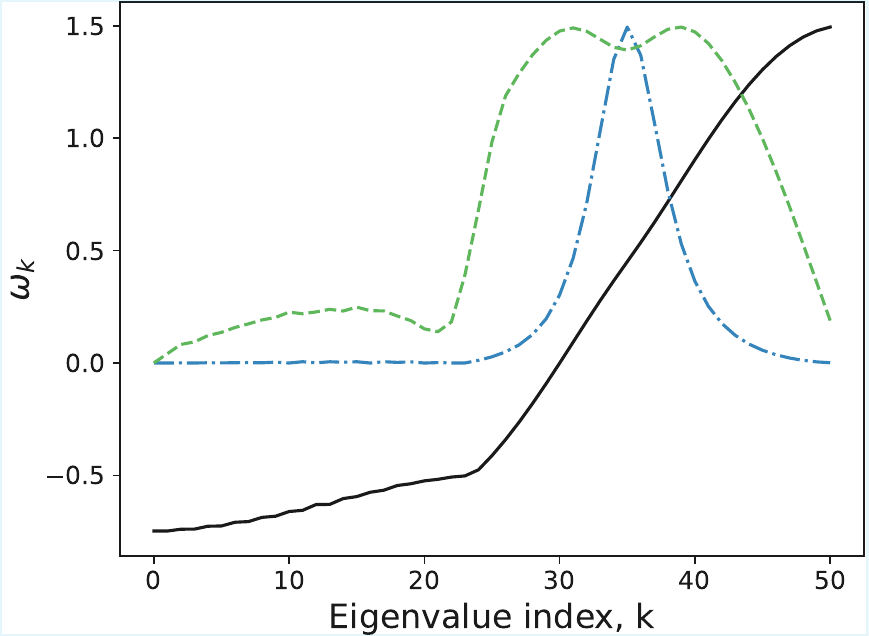}
        \put(15,60){\textbf{(c)}}
    \end{overpic}
    \end{minipage}\\[-0.74cm]
    \begin{minipage}[b]{.78\linewidth}
    \begin{overpic}[width=\linewidth]{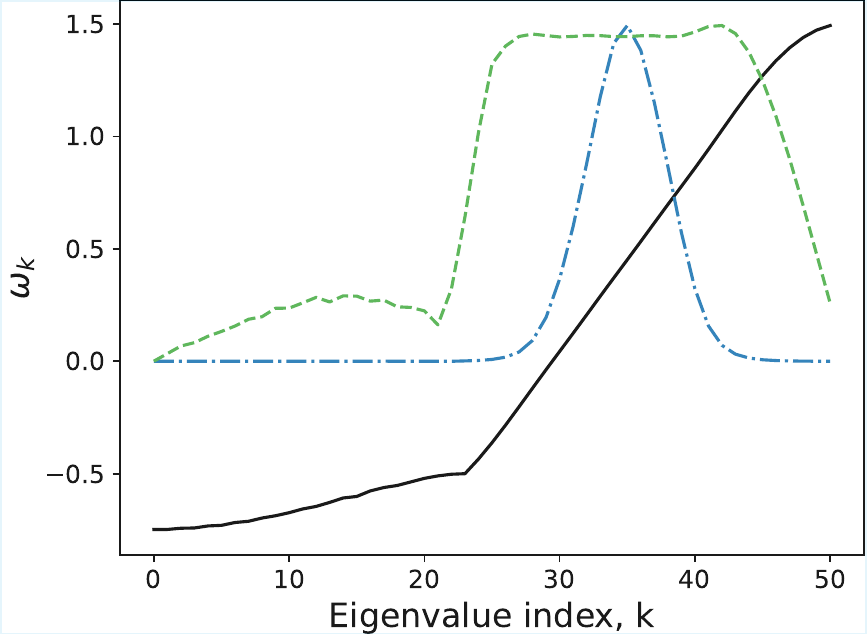}
        \put(15,60){\textbf{(d)}}
    \end{overpic}
    \end{minipage}
    \caption{Energy spectra (black), in units of $J_{1}$, for a 51-site next-nearest neighbour spin chain with different numbers of optimised couplings: (a) and (b) shows the cases where no couplings are optimised for both zero (a) and one (b) set of optimised on-site energies $h$, while (c) and (d) correspond to the cases with one and four pairs of optimised couplings respectively, in addition to $h$. The initial wave packet (blue, dash-dot), $v_{1k}^2$, and group velocity (green, dashed), $\frac{d \omega_k}{dk}$, are appropriately rescaled and overlaid on top of the spectra. All spectra are identical to those reported for $J_2 = \frac{1}{2}$ in Fig.~\ref{fig:dispersion_relation}, but ordered in increasing energy values.}
    \label{fig4}
\end{figure}

\begin{figure}
    \begin{minipage}[b]{.9\linewidth}
    \begin{overpic}[width=\linewidth]{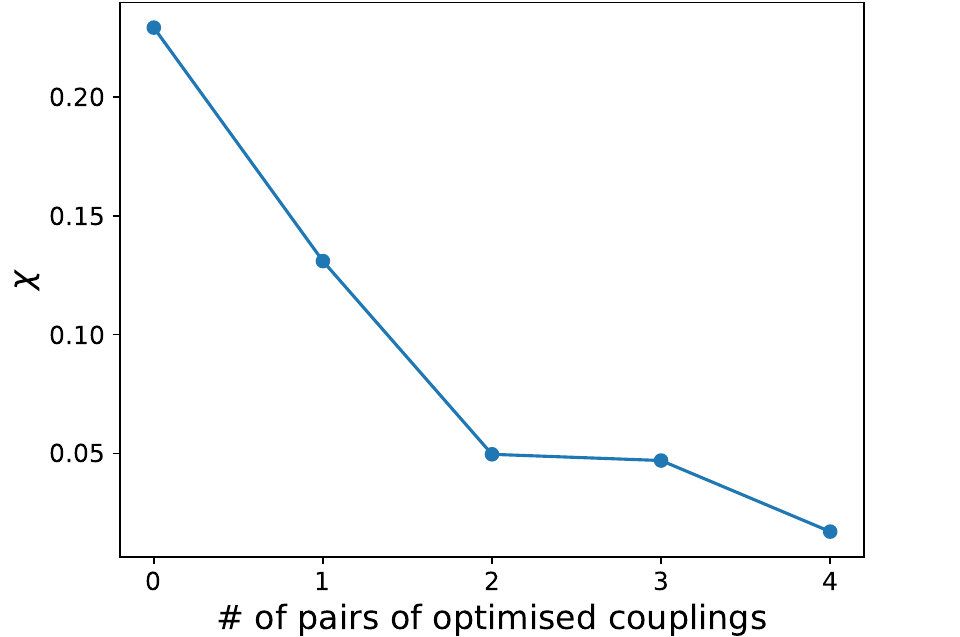}
        \put(75,50){\textbf{(a)}}
    \end{overpic}
    \end{minipage}\\[-0.7cm]
    \begin{minipage}[b]{.9\linewidth}
    \begin{overpic}[width=\linewidth]{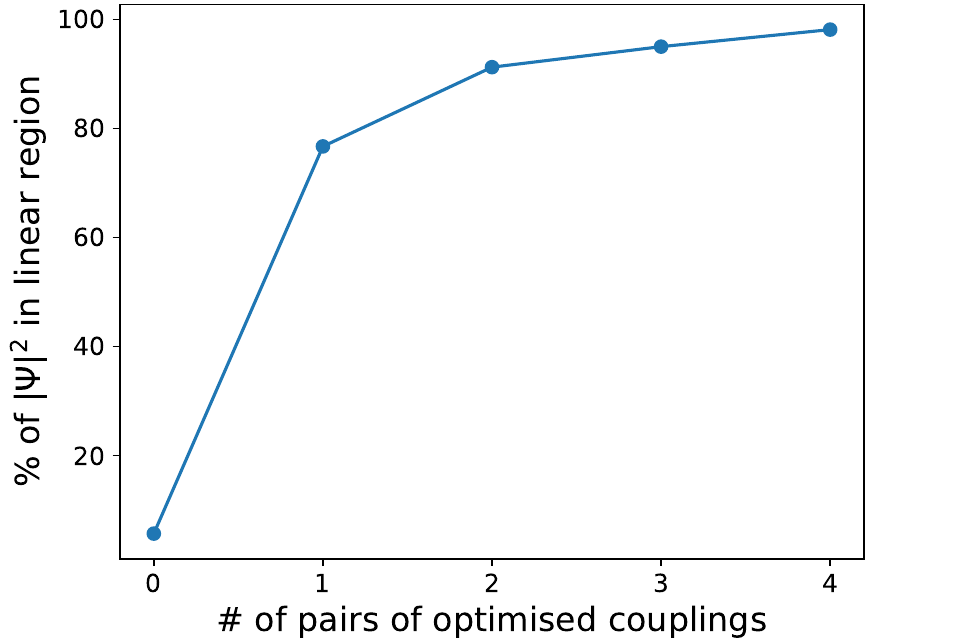}
        \put(75,50){\textbf{(b)}}
    \end{overpic}
    \end{minipage}
\caption{(a) Non-linearity and (b) percentage of the initial wave packet within a linear region, $\sum_k^{\{v_{1k}^2 \geq \epsilon\}}v_{1k}^2 $,  plotted against the number of pairs of optimised sites for $N=51$. The linear region is defined as the $k$-space where the second derivative in Eq.~\ref{eq:nonlinearity} is smaller than $10^{-3}$.}
\label{fig:energyspectr+h1+chivscouplings51site2NN-j-1-0.5}
\end{figure}

In order to provide a physical interpretation, and validate our intuitive picture, for the achievable fast and high-quality QST protocol, we analyse the impact of the optimisation procedure on the wave packet dynamics described by Eq.~\ref{eq:trans2}. In Figs.~\ref{fig4} we report the initial wave packet in $k$-space $v_{1k}^2$,  the wave packet group velocity $\frac{d \omega_k}{dk}$ and the dispersion relation $\omega_k$. The upper panels are with no optimisation on any $\alpha$'s and $\beta$'s, the left panel optimizing $h$; the lower panels are for one and four pairs of optimised couplings, in addition to $h$. In the absence of any optimised on-site and inter-site couplings (upper left panel), the average fidelity is only about 0.6 as, on the one hand, the wave packet (blue dot-dashed line) is centred around $k=\frac{N}{2}$ ($\omega_k=0$), and, on the other hand, the group velocity (green dashed line) is highly non-linear. Optimising the on-site energy $h$ (upper right panel) induces the shift of the peak of the wave packet to the inflection point of $\omega_k$, but its width is too large to result in dispersionless dynamics due to the non-linear dispersion relation giving a non-constant group velocity in the excited $k$-region.
From the lower panels - left with one set of optimised couplings and right with four sets - it can be seen both that the initial wave packet concentrates in the linear region and that the linear region of the dispersion relation increases by increasing the number of optimised couplings, as witnessed by the almost constant group velocity curve. We further pursue this analysis by determining which optimised couplings play a major role in the wave packet concentration and the dispersion relation linearisation, respectively. To do so we measure the degree of non-linearity of the dispersion relation around the inflection point via its second derivative:
\begin{equation}\label{eq:nonlinearity}
    \chi  = \sum_k \left |\frac{\partial^2 \omega_k}{\partial k^2}\right |, \qquad k \in \{v_{1k}^2  \geq \epsilon\},
\end{equation}
where $v_{1k}^2$ is the initial wave packet and $\epsilon$ is some small constant set to be 0.01. The chosen figure of merit in Eq.~\ref{eq:nonlinearity} quantifies how non-linear is the dispersion relation $\omega_k$ around the inflection point, considering a $k$-interval where the initial wave packet has a support greater than $\epsilon$. Clearly, the lower is $\chi$ the less non-linear is the dispersion relation, with $\chi=0$ the extreme case of a linear dispersion relation (in the chosen $k$-interval). 
Figures \ref{fig:energyspectr+h1+chivscouplings51site2NN-j-1-0.5}a and \ref{fig:energyspectr+h1+chivscouplings51site2NN-j-1-0.5}b show, respectively, the degree of non-linearity $\chi$ and the localisation of the wave packet in the linear region of the dispersion relation for increasing pairs of optimised NN and NNN couplings. It can clearly be seen that, for no optimised couplings (but with $h$ optimised), only a very small percentage of the wave packet is contained in the linear region and that the first set of couplings $\{\alpha_1,\beta_1\}$ accounts for the largest concentration effect. This is expected, as the excitation is placed on the first site and its couplings to the chain determine the width of the wave packet in $k$-space. Indeed, for $\{\alpha_1,\beta_1\}\rightarrow 0$, the width of the wave packet tends to zero as well, which is the coupling scheme exploited in Rabi-like QST dynamics~\cite{wojcikUnmodulatedSpinChains2005}.

Next we explore the effect of varying the NNN coupling $J_2$, by optimizing he end couplings for the 51-site spin chain for different values of  $J_2$.
Table~\ref{tab:poitable} shows that both optimised on-site energies and four sets of inter-site couplings $\{\alpha_i,\beta_i\}$ can be found such that the average fidelity is always above $99.8\%$ for all investigated $J_2$ values. The table also shows a column with the predicted value of the on-site energy, Eq.~\ref{eq_hvalues}: the closeness of the predicted and optimised values on the one side confirms the validity of the approximations reported in Sec.~\ref{sec2} for this size chain, and on the other highlights the efficacy of the genetic algorithm in optimising to such value. 

\begin{figure}
    \centering
\includegraphics[width=\linewidth]{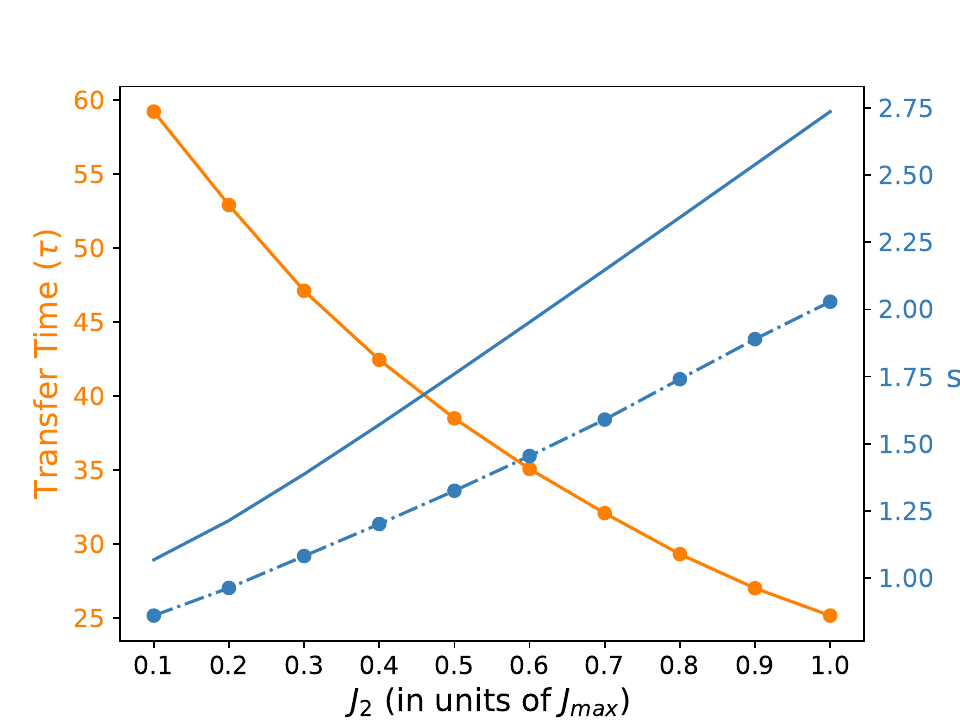}
    \caption{Transfer time (orange), in units of ($t \cdot J_{max}$), and ideal speed-up $s$ (blue, solid) from Eq \ref{eq:speedup} for four optimised sites over a nearest-neighbour chain (dash-dot) against $J_2$ for a 51-site spin chain. The transfer time clearly decreases by increasing $J_2$ with a speed-up $s$ that for $J_2=1$ is slightly larger than 2. 
}
    \label{fig:time}
\end{figure}

The speed-up in transfer time $\tau$ when increasing $J_2$ is shown in Figure \ref{fig:time}, where it closely follows the expected speed-up predicted in Eq. \ref{eq:speedup}. The widening gap between the numerical results and the approximation occurs due to a necessity for weaker end couplings to achieve high fidelity transfer.
\begin{table}[h!]
\centering

\renewcommand{\arraystretch}{1.3}
\begin{tabular}{|l|c|c|c|c|c|}
\hline
    \textbf{$J_2$} & $h_{p}$ & $h_o$ & \(\langle F(\tau)\rangle\) & $\tau$\\
    \hline
    0.1  & 0.4781 & 0.5328 & 0.9993 & 59.23 \\
    \hline
    0.2  & 0.6909 & 0.7029 & 0.9991 & 52.91\\
    \hline
    0.3  & 0.7932 & 0.7981 & 0.9993 & 47.11\\
    \hline
    0.4  & 0.8519 & 0.8988 & 0.9994 & 42.44\\
    \hline
    0.5  & 0.8896 & 0.8891 & 0.9993 & 38.48\\
    \hline
    0.6  & 0.9159 & 0.9394 & 0.9993 & 35.06\\
    \hline
    0.7  & 0.9352 & 0.9299 & 0.9990 & 32.07\\
    \hline
    0.8  & 0.9500 & 0.9781 & 0.9991 & 29.3\\
    \hline
    0.9  & 0.9616 & 0.9615 & 0.9989 & 26.99\\
    \hline
    1    & 0.9710 & 0.9593 & 0.9987 & 25.14\\
    \hline
    \end{tabular}
    \caption{Value of optimised $h$ ($h_o$), and predicted $h$ ($h_p$) for $N$ = 51  with the optimised values obtained for optimising all couplings on four external sites at each end of the chain as well as $h$.}
    \label{tab:poitable}
\end{table}

Finally, the highest fidelity optimisation scheme, i.e., optimising the couplings of four sites as well as the sender/receiver on-site energies, was applied to larger spin chains up to $N=500$. Table \ref{tab:differentN} presents some key features from the optimised spin chain dynamics, showing that the average fidelity decreases very slowly, of the order of $10^{-3}$, for the longest investigated chain. We conjecture that, also for arbitrarily large $N$, the average fidelity stays above $99\%$ as in the NN case~\cite{TA2012}, eventually by optimising a few additional inter-site couplings. In Fig.~\ref{fig:logNvstimeandcoupling} we report the scaling behaviour of the transfer time and the end site inter-spin couplings $\{\alpha_1,\beta_1\}$ as a function of $N$. The transfer time is expected to follow ballistic dynamics in the bulk, with a delay $\Delta t$ due to the rescaled end couplings, i.e. $\tau=\frac{N}{v_g}+\Delta t$, where $v_g=s=1.76$ as reported at the end of Sec.~\ref{sec2}. The orange line shows that the fitting curve $\Delta t \simeq 2.26 N^{0.363}$ implies a time delay polynomial with $N$ which is consistent with the scaling law of the weighted average of $\{\alpha_1,\beta_1\}$, $w=\frac{\alpha_1 J_1+\beta_1 J_2}{4(J_1+J_2)}$, which, in turn, is best fitted by the following scaling law $w=0.629 N^{-0.265}$. In fact, the closeness of these two scaling exponents is mainly due to boundary effects, where the modified end-couplings slow down the injection of the wave packet into the bulk of the chain. As the excitation is placed on the first site, it has two decay channels, $J_1$ and $J_2$, which transfer the excitation to the NN and the NNN site on a time scale respectively proportional to their inverse couplings, which motivates the choice of the weighted average $w$ in the scaling analysis.

\begin{figure}[h!]
\centering
\small
\renewcommand{\arraystretch}{1.1}

\begin{minipage}[c]{0.45\textwidth}
    \centering
    \begin{tabular}{|l|c|c|c|c|c|}
    \hline
    \textbf{N} & \(\lvert f_1^N(\tau)\rvert^2\)$_{\%}$ & \(\langle F(\tau)\rangle\) & $\tau$ & \(h\) & $w$ \\ \hline
    20   &   $99.88\%$    &     0.9996  &  18.06     &    0.8889 & 0.2782  \\ \hline
    50   &  $99.80\%$     &   0.9993    &  37.73     &  0.8999 & 0.2268    \\ \hline
    100  &     $99.73\%$  &    0.9991   &   68.89    &  0.8899 & 0.1880   \\ \hline
    150 & $99.71\%$ &  0.9990 & 99.40 & 0.8904 & 0.1681 \\ \hline
    200  & $99.67\%$& 0.9989& 129.09&    0.8836 & 0.1553 \\ \hline
    250 & $99.66\%$ & 0.9988 & 159.08 & 0.8953 & 0.1427 \\ \hline
    500  &  $99.51\%$     &  0.9984     &    305.44   &  0.8947 & 0.1200 \\ \hline
    \end{tabular}
    \captionof{table}{Optimised system parameters for various chain lengths \(N\), including on-site field \(h\). Maximal fidelity scores are attained within time window \(t \cdot J_{\text{max}} \in [0, N]\).}
    \label{tab:differentN}
\end{minipage}
\hfill
\begin{minipage}[c]{0.49\textwidth}
    \centering
    \includegraphics[width=\linewidth]{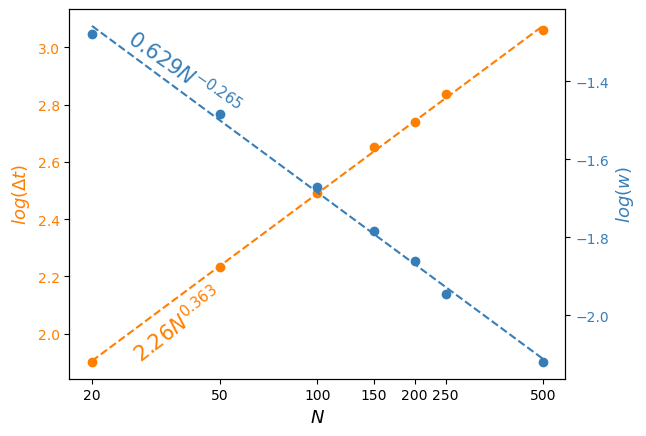}
    \caption{Scaling laws of the time delay $\Delta t$ from ideal transfer time (orange) and of the weighted sum $w$ of the coupling strengths $\{\alpha_1,\beta_1\}$ (blue) against $N$. Fits to the data are shown with dashed lines.}
    \label{fig:logNvstimeandcoupling}
\end{minipage}
\end{figure}

\subsection{Comparison with NNN PST Scheme}
Here we compare our coupling scheme with a known solution for achieving PST within NNN XY spin chains~\cite{Christandl2}. Similar to the nearest-neighbour model, the NNN PST coupling scheme requires a parabolic configuration of both the NN and NNN couplings, with the coupling differences between the first and middle sites increasing as a function of \( N \). In addition, a non-negligible and relatively large parabolic magnetic field profile, throughout the entirety of the chain, is required to enable PST. The authors of \cite{Christandl2} note that, under this scheme, one can either let the transfer time $\tau$ scale proportionally with \( N \), in an attempt to minimise the very large coupling energies in the bulk (which increases with N), or use the PST coupling scheme mainly for short-haul transfer.
\begin{figure}[h!]
    \centering

    \begin{subfigure}[t]{0.48\textwidth}
        \centering
        \includegraphics[width=\linewidth]{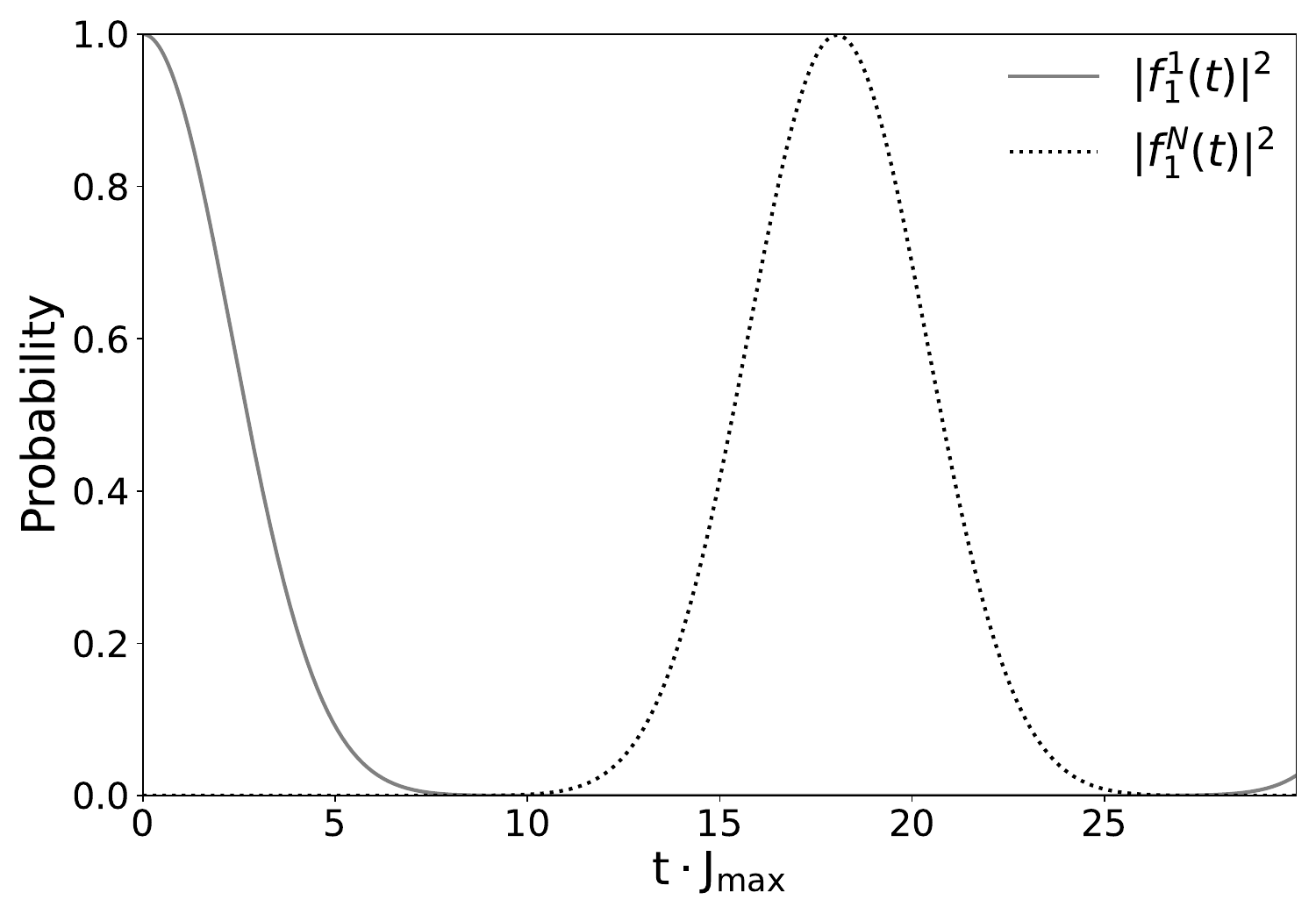}
        \caption{Transfer dynamics with the coupling scheme generated by the genetic algorithm.}
        \label{fig:ga-dynamics}
    \end{subfigure}
    \hspace{-0.0cm}
    \begin{subfigure}[t]{0.45\textwidth}
        \centering
        \includegraphics[width=\linewidth]{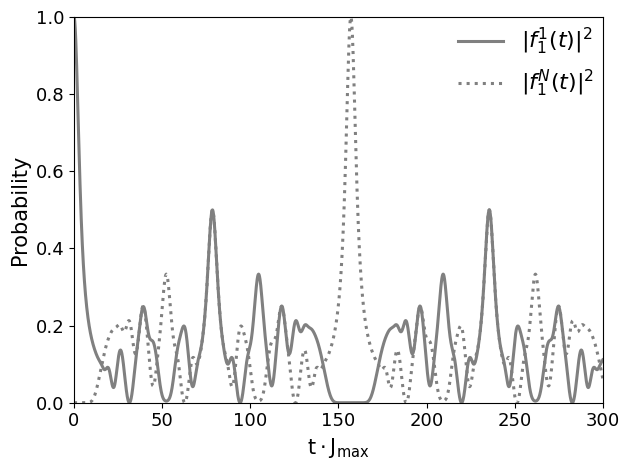}
        \caption{Transfer dynamics with the coupling scheme introduced in Ref.~\cite{Christandl2}, with $\frac{\alpha_{\tiny{\cite{Christandl2}}}}{\beta_{\tiny{\cite{Christandl2}}}}= 1$ (in their notation).}
        \label{fig:christandl-dynamics}
    \end{subfigure}

    \caption{Comparison of QST coupling schemes for an $N = 20$ NNN chain. The dotted (continuous) line represent the probability $\left|f_1^N(t)\right|^2$ ($\left|f_1^1(t)\right|^2$) of the excitation to be found on the receiver (sender) at time $t$.} 
    \label{fig:side-by-side}
\end{figure}

In Fig.~\ref{fig:side-by-side} we compare the QST dynamics for $N=20$ of our coupling scheme (left) and the known PST result (right). In the left panel, it is seen how the probability of the excitation to reside on the last (first) site is monotonically increasing (decreasing), signalling a quasi-dispersionless ballistic dynamics. On the other hand, in the right panel, the occupation probabilities on the first and last site for the PST scheme are widely oscillating, a signature that PST is achieved via carefully engineered couplings yielding fully constructive interference of the wave packet components on the last site at a predictable time. It is also worth mentioning that the ballistic QST protocol is completed in a much shorter time than for the comparable PST coupling scheme. This property of our proposed coupling scheme may prove very useful in spin-$\frac{1}{2}$ experimental implementations with short coherence times.
Furthermore, as shown in Fig.~\ref{fig:side-by-side}, the transfer dynamics of the PST coupling scheme
exhibit sharp spikes implying a narrow high-quality read-out time window of the received quantum state.
In contrast, the ballistic dynamics induced by our proposed coupling scheme
allows for a wider time interval during which the QST fidelity remains high. This feature is relevant when timing errors may occur, e.g. in recovering the transferred state and in synchronizing injection of different excitations or of other local operations \cite{D2007,D2009,R2011,Estarellas_2017,Estarellas_2020}; the notion of evaluating the quality of transfer dynamics based on the temporal width of the fidelity peaks
is discussed in detail within Refs.~\cite{Kay1,kay2025optimisingperfectquantumstate}. 

\subsection{Potential experimental implementations}
The XY NNN Hamiltonian has been successfully simulated in a variety of physical platforms \cite{Shea97,Nikos2002,Toskovic2016,Shi2024}. As shown in Table~\ref{tab:hardware}, the state transfer protocols here discussed can, in principle, be implemented on several types of quantum computing hardware that exhibit next-nearest-neighbour interactions, making them suitable for information transfer tasks. These platforms offer a trade-off between interaction strength, transfer time, and coherence time, which must be balanced for practical implementations.

\begin{table}[h!]
\small
\centering
\begin{tabular}{|l|c|c|c|}
\hline
\hspace{1.cm}\textbf{Hardware} & \(\mathbf{{J_{max}/\hbar}}\) & \(\mathbf{\tau}\) &\(\mathbf{T_{2}}\) \\ \hline
Nanomechanical Lattices \cite{Tian2020} & 1 MHz & 40 $\mu$s & 50~$m$s \\ \hline
Superconductors \cite{Xiang} & 10 GHz & 4 ns & 50~$\mu$s \\ \hline
Cold Atoms \cite{Cheng} & 1 MHz & 40~$\mu$s & $\sim$~seconds \\ \hline
Waveguide Arrays \cite{Sheremet2023} & 50 MHz & 1~$\mu$s & $\sim$~$\mu$s--$m$s \\ \hline
\end{tabular}
\caption{Comparison of different hardware in terms of operation time and coherence. $J_{max}$ corresponds to the typical energy scale estimated from the two-qubit gate operation time $\tau_2$ reported in the cited literature, and $\tau$ is the estimated time (in seconds) required for high-fidelity transfer for $N = 51$ ($\tau$ = 40/$J_{max}$)) which is within the dephasing time $T_2$.}
\label{tab:hardware}
\end{table}

\section{Conclusions}\label{conc}
We have presented a fast and high-fidelity QST protocol in long-range interaction spin-$\frac{1}{2}$ models. While we focused on the NNN model, the physical mechanism underlying the quasi-dispersionless wave packet dynamics can be readily extended to models with a longer interaction range. Identifying the linear region of the dispersion relation is key in order to determine the value of the on-site end energies such that the peak of the wave packet is centred at the inflection point. Subsequently, reducing the value of a few end couplings is sufficient to both concentrate the wave packet in the linear region and, simultaneously, increase the degree of linearity of the dispersion relation around the inflection point. In most of our work we have chosen the next-nearest-neighbour coupling $J_2$ to be half the value of the nearest-neighbour coupling, i.e. $J_2=\frac{J_1}{2}$ and have found, utilising the genetic algorithm as optimisation procedure, that modifying the couplings of just four sites on each end is sufficient to achieve an average fidelity above $99.8\%$ in chains as long as 500 sites. Interestingly, the proposed coupling scheme yields a high-fidelity QST time that is significantly lower than that with NNN PST couplings. Furthermore, as the wave packet travels with a speed close to the maximum group velocity for NNN interacting spin-$\frac{1}{2}$ models, we suppose that our achieved QST times are very close to those set by their relevant quantum speed limits.

It is also worthwhile to notice that, increasing the next-nearest neighbour coupling or, equivalently, including additional interaction terms for third- and fourth-nearest neighbour sites, results in a higher group velocity, but at the price of reducing the size of the linear region of the dispersion relation. This observation hints towards the fact that there is a threshold between fidelity and QST time, i.e., faster transfer times may come at the expense of a reduced fidelity, similarly to thresholds that are found in quantum clock theory~\cite{PhysRevLett.131.220201}. Finally, it would be interesting to explore potential advantages of using long-range interacting systems for enhancing quantum clock's precision~\cite{meierPrecisionNotLimited2025}, for the transfer of more than a single qubit~\cite{apollaroQuantumTransferInteracting2022} or for reducing the sensitivity with respect to timing readout errors~\cite{kay2025optimisingperfectquantumstate} or for energy transport~\cite{murphy2025ergotopytransportdimensionalspin}.

\section{Acknowledgements}
TJGA and IDA acknowledge funding from the Royal Society under the grant IES\textbackslash R3\textbackslash 243264 - International Exchanges 2024 Global Round 3.

\noindent C.C. Nelmes acknowledges support from EPSRC, grant number is EP\textbackslash W524657\textbackslash1.
\appendix
\bibliography{ref}

\end{document}